\documentstyle[twocolumn,aps]{revtex}
\begin{document}
\draft

\title{Critical Behavior in Black Hole Thermodynamics}

\author{Rong-Gen Cai}

\address{Center for Theoretical Physics,\\
 Seoul National University, Seoul 151-742, Korea}

\maketitle

\begin{abstract}
In this talk I  introduce the critical behavior occurring at the
extremal limit of black holes. The extremal limit of black holes 
is a critical point and  a  phase transition takes place from the
extremal black holes to their nonextremal counterparts. Some 
critical exponents satisfying the scaling laws are  obtained.  From
the scaling laws we introduce the concept of the effective dimension
of black holes and discuss the  relationship between the critical 
behavior and the statistical interpretation of black hole entropy. 

\end{abstract}

%\newpage

\section{Introduction}

Due to the celebrated works of Hawking \cite{Haw} and Bekenstein
\cite{Bek}, a black hole has a temperature proportional to the surface
gravity on the black hole horizon and an entropy proportional 
to the area of event horizon. This put the first law of black hole 
thermodynamics on  a solid fundament \cite{Bar}. Also this made people 
believe  that a black hole is a thermodynamic system.

The phase transition is an important phenomenon in the ordinary
thermodynamics. It is therefore natural to ask whether 
 there  is any phase  transition in the black hole thermodynamics.
 Davies \cite{Davies}  is the first 
person who  discusses  the critical behavior of Kerr-Newman black hole 
in terms of thermodynamics. For a Kerr-Newman black hole with mass
$M$, charge $Q$ and angular momentum $J=aM$, the black hole entropy
is
\begin{equation}
S=\frac{A}{4}=2\pi \left (M^2-\frac{1}{2}Q^2 +\sqrt{M^4-J^2-M^2Q^2}
           \right),
\end{equation}
and the Hawking temperature is
\begin{equation}
T=\frac{\kappa}{2\pi}=
   \frac{\sqrt{M^2-a^2-Q^2}}{2M^2-Q^2+2M\sqrt{M^2-a^2-Q^2}}.
\end{equation}
According to the formula $C_{J,Q}=(\partial M/\partial T)_{J,Q}$, 
the heat capacity is
\begin{equation}
C_{J,Q}=\frac{MTS^3}{\pi J^2 +\pi Q^4/4-T^2S^3}.
\end{equation}
For the Schwarzschild black hole, $C=-M/T <0$; For an extremal 
Kerr-Newman black hole, $C_{J,Q}\rightarrow 0^+$. Thus the heat 
capacity must diverge at certain points:
\begin{equation}
\left \{
\begin{array}{ll}
Q_c=\sqrt{3}M/2, & {\rm for~~ RN~~ black~~holes}\\
J_c=\sqrt{2\sqrt{3}-3}M^2, & {\rm for~~ Kerr~~ black~~ holes}
\end{array} \right.
\end{equation}
Based on the infinity discontinuity of the heat capacity,
Davies claimed that there are second-order phase transitions
in black hole thermodynamics. Many authors have investigated 
the critical points in the different black holes 
\cite{Hut,Soko,Lau,Muni}. Lousto \cite{Lousto}  claimed 
that the critical points of Davies satisfy the fourth law of
black hole thermodynamics: scaling laws, from which  he assigned
the effective spatial dimension of black holes to be  two, 
and thought that 
his result is in complete agreement with the membrane picture
of black holes \cite{Thorne}. However, he found later that 
his calculation has some errors. Thus his two-dimensional effective
model of black holes becomes invalid.

On the other hand, some people think that there exist a critical 
point at the extremal limit of black holes and a second-order 
phase transition takes place from an extremal black hole to 
its nonextremal counterpart \cite{Curir}. Kaburaki \cite{Kab}
found that
the critical point at extremal limit also obeys the scaling 
laws by investigating the thermal equilibrium fluctuations of 
Kerr-Newman black hole in the micro-canonical ensemble. The present 
author and collaborators have investigated in detail the critical 
behavior in the BTZ black holes \cite{Cai1}, dilaton black holes
\cite{Cai2}, and black $p$-branes \cite{Cai3}, respectively,
and obtained some
interesting results. This article is devoted to showing these 
results in the
nondilatonic black $p$-branes and  arbitrary dimensional dilatonic
black holes.

The organization of this paper is as follows. In next
section we will discuss the critical behavior in the nondilatonic 
black $p$-branes and calculate the relevant critical exponents. In  
Sec.~III we investigate the case of arbitrary dimensional dilatonic 
black holes. The conclusion and discussion are presented in Sec.~IV.

\section{Critical behavior for  nondilatonic black $p$-branes}

The nondilatonic black $p$-branes we will consider 
come from the following 
action \cite{Gibbons}
\begin{equation}
\label{action1}
S_{d+p}=\frac{1}{16\pi}\int d^{(d+p)}x\sqrt{-g}
\left[R-\frac{2}{(d-2)!}F^2_{d-2}\right],
\end{equation}
where $R$ is the scalar curvature and $F_{d-2}$ denotes the
($d-2$)-form anti
symmetric tensor field. Performing the
 double-dimensional reduction by $p$ dimensions, one has the
 dilatonic $d$-dimensional action:
\begin{equation}
\label{action2}
S_d=\frac{1}{16\pi}\int d^d\sqrt{-g}\left[R-2(\nabla \phi)^2-
\frac{2}{(d-2)!}e^{-2a\phi}F^2_{d-2}\right],
\end{equation}
where $\phi$ is the dilaton field, and the constant $a$ is
\begin{equation}
\label{parama}
a=\frac{(d-3)\sqrt{2p}}{\sqrt{(d-2)(d+p-2)}}.
\end{equation}
The magnetically charged black holes in the action (\ref{action2}) are 
\begin{eqnarray}
\label{solution}
ds^2_d&=&\left[1-\left(\frac{r_+}{r}\right)^{d-3}\right]
\left[1-\left(\frac{r_-}{r}\right)^{d-3}\right]^{1-(d-3)b}dt^2
\nonumber\\
&+&\left[1-\left(\frac{r_+}{r}\right)^{d-3}\right]^{-1}
\left[1-\left(\frac{r_-}{r}\right)^{d-3}\right]^{b-1}dr^2
\nonumber\\
&+&r^2\left[1-\left(\frac{r_-}{r}\right)^{d-3}\right]^{b}
d\Omega^2_{d-2}, \nonumber \\
e^{a\phi}&=&\left[1-\left(\frac{r_-}{r}\right)^{d-3}
\right]^{-(d-3)b/2}, \nonumber \\
F_{d-2}&=&Q\varepsilon _{d-2},
\end{eqnarray}
where $\varepsilon _{d-2}$ is the volume form on the unit
 ($d-2$)-sphere, the constant $b$ is
 \begin{equation}
 \label{paramb}
 b=2p/(d-2)(p+1),
 \end{equation}
 and the charge $Q$ is related to $r_{\pm}$ by
 \begin{equation}
 Q^2=\frac{(d-3)(d+p-2)}{2(p+1)}(r_+r_-)^{d-3}.
 \end{equation}
 Thus, one has the non-dilatonic black $p$-brane solutions in
 the action (\ref{action1}) \cite{Gibbons} 
 \begin{equation}
 \label{brane}
 ds^2_{d+p}=e^{2m\phi}dy^idy^i + e^{2n\phi}ds^2_d,
 \end{equation}
 where $i=1,2,\cdots,p$, and
 \begin{equation}
 m=-\frac{\sqrt{2(d-2)}}{\sqrt{p(d+p-2)}}, \ \ n=-\frac{mp}{d-2}.
 \end{equation}
 The Hawking temperature and the
 Bekenstein-Hawking
 entropy per unit volume of $p$-branes for the black $p$-branes
 (\ref{brane})
  are easily obtained:
  \begin{eqnarray}
  \label{temp1}
  && T=\frac{d-3}{4\pi r_+}\left
  [1-\left(\frac{r_-}{r_+}\right)^{d-3}
  \right]^{1/(p+1)},\\
  \label{entropy1}
  && S=\frac{\Omega _{d-2}}{4}r_+^{d-2}\left[1-\left(\frac{r_-}{r_+}
  \right)^{d-3}\right]^{p/(p+1)},
  \end{eqnarray}
  where $\Omega _{d-2}$ is the volume of the unit ($d-2$)-sphere.
   The ADM mass per
 unit volume of $p$-branes is found to be
 \begin{equation}
 \label{adm}
 M=\frac{\Omega _{d-2}}{16\pi}\left[(d-2)r_+^{d-3}+\frac{d-2-p(d-4)}
 {p+1}r_-^{d-3}\right],
 \end{equation}
 which satisfies the first law of thermodynamics,
 \begin{equation}
 \label{first}
 dM=TdS+\Phi dQ,
 \end{equation}
 where $\Phi=\Omega _{d-2}Q/[4\pi (d-3)r_+^{d-3}]$ is the chemical
  potential corresponding to the conservative charge $Q$. 
  The heat
	capacity per unit volume of $p$-branes is
	\begin{eqnarray}
	C_Q&=&-\frac{\Omega _{d-2}r_+}{4}\frac{\left[1-
	\left(\frac{r_-}{r_+}\right)^{d-3}\right]^{p/(p+1)}}
{\left[1-\frac{p+2d-5}{p+1}\left(\frac{r_-}{r_+}\right)^{d-3}
\right]} \nonumber\\
&\times&
\left[(d-2)r_+^{d-3}-
\frac{d-2-p(d-4)}{p+1}r_-^{d-3}\right].
\end{eqnarray}
When the extremal limit ($r_-=r_+$) is approached, the temperature,
 entropy, and the heat capacity approach zero. When
 $1-(p+2d-5)(r_-/r_+)^{d-3}/(p+1)=0$, the heat capacity diverges,
 which  corresponds to the critical point of Davies in Kerr-Newman
black holes \cite{Davies}.

In a self-gravitating thermodynamic system, the thermodynamical ensembles
are not equivalent generally. To study the critical behavior in black
hole thermodynamics, it is reasonable to choose the micro-canonical
ensemble \cite{Kab,Cai1,Cai2,Cai3}, in which the thermodynamical
potential function is the 
entropy of system. Rewriting (\ref{first}) we have 
\begin{equation}
\label{first2}
dS=\beta dM -\varphi dQ,
\end{equation}
where $\beta =T^{-1}$ and $\varphi=\beta \Phi$. From (\ref{first2})
it follows
that the intrinsic variables are $\{ M,Q\}$ and the conjugate variables
$\{ \beta, -\varphi\}$.Thus the eigenvalues
corresponding to the fluctuation modes $\beta$ and $\varphi$ are
\begin{eqnarray}
&&\lambda _m=\left(\frac{\partial M}{\partial \beta}\right)_Q=
-T^2C_Q,\\
&&\lambda _q=-\left(\frac{\partial Q}{\partial \varphi}\right)_M=
-TK_M,
\end{eqnarray}
respectively, where
\begin{eqnarray}
\label{km}
K_M&\equiv & \beta\left(\frac{\partial Q}{\partial \varphi}
\right)_M \nonumber\\
&=&\frac{4\pi (d-3)r_+^{d-3}}{\Omega _{d-2}}\left[1-\frac{d-2
-p(d-4)}{(p+1)(d-2)}
\left(\frac{r_-}{r_+}\right)^{d-3}\right] \nonumber\\
&&\left \{\left[1+\frac{d-2-p(d-4)}{p+1)(d-2)}\left(\frac{r_-}
{r_+}\right)^{d-3} \right] \right. \nonumber\\
&+& \frac{2}{(d-3)}\left(\frac{r_-}{r_+}\right)^{d-3}\left[
\frac{d-3}{p+1}-
\frac{d-2-p(d-4)}{(p+1)(d-2)} \right. \nonumber\\
&\times&\left. \left(1-\frac{p+d-2}{p+1}\left(\frac{r_-}{r_+}
\right)^{d-3}\right)\right] \nonumber\\
&\times& \left. \left[1-\left(\frac{r_-}{r_+}\right)^{d-3}
\right]^{-1}\right\}^{-1}.
\end{eqnarray}
Obviously, the two eigenvalues approach zero as the extremal 
limit is approached. Hence some second moments must diverge 
at the extremal limit \cite{Cai3}. As in the ordinary 
thermodynamics, the  divergence of
second moments means that the extremal limit is a critical point
and a second-order phase transition
takes place from the extremal to nonextremal black $p$-branes.
As is well known, the extremal black $p$-braes are very different
from the nonextremal in many aspects, such as the thermodynamic
description  and geometric structures \cite{Cai1}. In particular,
it has been shown that some  extremal black $p$-branes are
 supersymmetric and the supersymmetry is absent for the
 nonextremal  black $p$-branes. So the occurrence of phase 
 transition  are consistent with
 the changes of symmetry. The extremal and nonextremal black
 $p$-branes  are two different phases. The extremal black $p$-branes
 are in the disordered phase and the nonextremal black
 $p$-branes
 in the ordered phase. The order parameters
 of the phase transition can be defined as the differences of
 the conjugate variables between the two 
 phases \cite{Kab,Cai1,Cai2,Cai3}, such
 as  $\eta _{\varphi}=
 \varphi _+-\varphi _-$ can be regarded as the order
 parameters 
  of  black $p$-branes, where the suffixes ``$+$'' and ``$-$''
 mean that the quantity is taken at the $r_+$ and $r_-$, 
respectively. The second-order derivatives of entropy with respect 
to the intrinsic variables are the inverse eigenvalues,
\begin{eqnarray}
\zeta_m&\equiv &\left (\frac{\partial ^2S}{\partial M^2}\right)_Q=
\lambda _{m}^{-1}=-\frac{\beta ^2}{C_Q},\\
\zeta_q&\equiv&\left (\frac{\partial ^2 S}{\partial Q^2}\right)_M=
\lambda _{q}^{-1}=-\frac{\beta }{K_M}.
\end{eqnarray}
 Correspondingly, we can define the critical exponents of these 
 quantities as follows,
\begin{eqnarray}
\zeta _m&\sim&\varepsilon _M^{-\alpha}\ \ \ 
({\rm for\ Q\ fixed}),\nonumber\\
         &\sim&\varepsilon _Q^{-\psi}\ \ \ ({\rm for\ M\ fixed}),\\
\zeta_q & \sim & \varepsilon ^{-\gamma}_M \ \ \  
({\rm for\ Q\ fixed }),      \nonumber \\
          & \sim &\varepsilon ^{-\sigma}_Q \ \ \ ({\rm for\ M\ fixed}),\\
\eta _{\varphi}&\sim &\varepsilon ^{\beta}_M \ \ \ 
({\rm for\ Q\ fixed}),\nonumber\\
      &\sim &\varepsilon ^{\delta ^{-1}}_Q \ \ \ ({\rm for\ M\ fixed}),
\end{eqnarray}
where $\varepsilon _M$ and $\varepsilon _Q$ represent the 
infinitesimal deviations of $M$ and $Q$ from their limit values. 
These critical exponents are found to be
\begin{equation}
\alpha=\psi=\gamma=\sigma=\frac{p+2}{p+1}, \ \
 \beta=\delta ^{-1}=-\frac{1}{p+1}.
\end{equation}
 The critical exponents $\beta $ and $\delta ^{-1}$ are negative, 
which shows the fact that the order parameter $\eta _{\varphi}$ 
diverges at the extremal limit. This is because the critical 
temperature is  zero in this phase transition. It is easy to 
check that these critical
  exponents satisfy the scaling laws of the ``first kind,''
\begin{equation}
\label{scaling1}
\alpha +2\beta +\gamma =2,\ \
\beta(\delta-1)=\gamma, \ \
\psi (\beta +\gamma)=\alpha.
\end{equation}
That scaling laws (\ref{scaling1}) hold for the black $p$-branes 
is related to 
the fact that the black $p$-brane entropy (\ref{entropy1}) is a 
homogeneous function, satisfying 
\begin{equation}
S(\lambda M, \lambda Q)=\lambda ^{(d-2)/(d-3) }S(M,Q),
\end{equation}
where $\lambda$ is a positive constant.

On the other hand, 
in an ordinary thermodynamic system, an important physical 
quantity related to phase transitions is the two-point 
correlation function, which has  
generally the form for a large distance,
\begin{equation}
G(r)\sim \frac{\exp (-r/\xi)}{r^{\bar{d}-2+\eta}},
\end{equation}
where $\eta$ is the Fisher's exponent, $\bar{d}$ is the effective
 spatial dimension of the system under consideration, and $\xi$ is
  the correlation length and diverges at the critical point. 
Similarly, the critical exponents of 
  the correlation 
  length for black $p$-branes can be defined as:
\begin{eqnarray}
\xi &\sim & \varepsilon ^{-\nu}_M \ \ \ ({\rm for\ Q\ fixed}),
\nonumber\\
    &\sim &\varepsilon ^{-\mu}_Q \ \ \ ({\rm for\ M\ fixed}).
\end{eqnarray}
Combining with those in Eq.~(\ref{scaling1}), these critical 
exponents form the scaling laws of the ``second kind''
\begin{equation}
\label{scaling2}
\nu (2-\eta )=\gamma, \ \ \nu \bar{d}=2-\alpha, \ \ 
\mu (\beta +\gamma)=\nu.
\end{equation}
Because of the absence of quantum theory of gravity, we have not 
yet the correlation function of quantum black holes. 
 Here we use 
the correlation function of scalar fields on the background of 
these black $p$-branes to mimic the one of black $p$-branes (from
 the obtained result below, it seems an appropriate approach to 
study the critical behavior of black holes at the present time).
 From the work of Traschen \cite{Traschen} who studied the behavior of 
 scalar fields on the background of Reissner-Nordstr\"{o}m black
  holes, it is found that the inverse surface gravity of the black
 hole  plays the role of the correlation length of scalar 
fields. For the black $p$-branes, this conclusion holds as 
well. With the help of the surface gravity of black 
$p$-branes (\ref{temp1}), we obtain
\begin{equation}
\label{munu}
\nu=\mu=\frac{1}{p+1}.
\end{equation}
 Substituting (\ref{munu}) into (\ref{scaling2}), we find
\begin{equation}
\eta=-p,\ \ \bar{d}=p.
\end{equation}
When $p=0$, the black $p$-branes (\ref{brane}) reduce to the 
non-dilatonic
 $d$-dimensional black holes. In this case,  
a similar calculation gives
\begin{eqnarray}
&&\alpha=\psi=\gamma=\sigma= 3/2,\ \ \beta=\delta ^{-1}=-1/2,
\nonumber\\
&& \eta=-1, \ \ \bar{d}=1,
\end{eqnarray}
from which it is easy to see that these critical exponents are 
independent of the dimensionality of spacetime and parameters of 
black  holes. These 
critical exponents are exactly the same as those of three 
dimensional BTZ black holes \cite{Cai1}. 
Recall the fact
that the BTZ black holes are also exact non-dilatonic black hole 
solutions in string theory, we find that these critical 
exponents are universal for non-dilatonic black holes, an 
important feature of critical behavior in the
non-dilatonic black holes.   For the dilatonic black holes 
with the coupling constant $a$ obeying (\ref{parama}), we find
that the effective spatial dimension is also $p$ 
(it is one for $p$=0). For a general $a$, the scaling
laws  still hold, but these 
critical exponents and effective dimension will depend on the 
coupling constant $a$ and the dimension $d$ of spacetime. In the next 
section, for the sake of generality, 
we will discuss case of an arbitrary dimensional dilatonic black
holes.

\section{Critical behavior for arbitrary dimensional dilatonic
black holes}

If the coupling constant $a$ between the dilaton and $(d-2)$-form is not
related to the parameter $p$ in the action (\ref{action2}), instead of 
$b$ in (\ref{paramb}), the parameter $b$ in the magnetically charged 
black hole solutions (\ref{solution}) should be 
\begin{equation}
b=2a^2(d-2)/[(d-3)(2(d-3)+a^2(d-2)].
\end{equation}
And the charge $Q$ becomes
\begin{equation}
Q^2=\frac{(d-2)(d-3)^2}{2(d-3)+(d-2)a^2}(r_+r_-)^{d-3}.
\end{equation}
The ADM mass of the solution is 
\begin{equation}
M=\frac{\Omega
_{d-2}(d-2)}{16\pi}\left[r_+^{d-3}+\frac{2(d-3)-(d-2)a^2}
{2(d-3)+(d-2)a^2}r_-^{d-3}\right].
\end{equation}
The Hawking temperature and Bekenstein-Hawking entropy of the holes are
\begin{eqnarray}
T&=&\frac{d-3}{4\pi
r_+}\left[1-\left(\frac{r_-}{r_+}\right)^{d-3}\right]^{1-(d-2)b/2},\\
S&=&\frac{\Omega _{d-2}r_+^{d-2}}{4}\left[1-\left(\frac{r_-}
{r_+}\right)^{d-3}
\right]^{(d-2)b/2},
\end{eqnarray}
respectively.   In this  case, at the extremal limit ($r_-=r_+)$,
the entropy always vanishes because of $b>0$. But the behavior of
Hawking temperature strongly depends on the parameter $a$. When
$b=2/(d-2)$, the limiting temperature is finite. When $b>2/(d-2)$, 
the Hawking temperature diverges at the extremal limit. The Hawking 
temperature is zero as $b<2/(d-2)$. The behavior of thermodynamics is
a general characteristic of dilatonic black hole thermodynamics.

Repeating the calculations in the previous section, we have 
the heat capacity
\begin{eqnarray}
C_Q &=&-\frac{(d-2)\Omega _{d-2}r_+^{d-2}}{4}
\left[1-\left(\frac{r_-}{r_+}\right)^{d-3}\right]^{(d-2)b/2}
        \nonumber \\
    &\times&		  
  \frac{ \left[ 1-\frac{2(d-3)-
  (d-2)a^2}{2(d-3)+(d-2)a^2}\left(\frac{r_-}{r_+}\right)^{d-3}
  \right]}
      {\left\{1-\left[2(d-3)+1-\frac{2(d-2)^2a^2}{2(d-3)+
		(d-2)a^2}\right]
        \left(\frac{r_-}{r_+}\right)^{d-3}\right\} }.
\end{eqnarray}
and $k_M$ in Eq.~(\ref{km}) is
\begin{eqnarray}
K_M &\equiv &\beta \left(\frac{\partial Q}{\partial \varphi}\right)_M
\nonumber\\
&=&\frac{4\pi (d-3)}{\Omega _{d-2}}\frac{\left[1-\frac{2(d-3)-(d-2)a^2}
{2(d-3)+(d-2)a^2}\left(\frac{r_-}{r_+}\right)^{d-3}\right]}
{\left[1+\frac{2(d-3)-(d-2)a^2}
{2(d-3)+(d-2)a^2}\left(\frac{r_-}{r_+}\right)^{d-3}\right]}
\nonumber \\
&\times &
\left\{\frac{1}{r^{d-3}}+\frac{2(d-3)^2-(d-2)a^2}{2(d-3)+(d-2)a^2}
\right.\nonumber\\
&\times& \left.
\frac{2r_-^{d-3}}{(d-3)r_+^{2(d-3)}}\left[1-\left(\frac{r_-}{r_+}
  \right)^{d-3}
\right]\right\}.
\end{eqnarray}
Thus we have the nonvanishing second moments in the micro-canonical
ensemble,
\begin{eqnarray}
&& \langle \delta \beta \delta \beta \rangle =-k_B\frac{\beta ^2}{C_Q},\ \
 \langle \delta \varphi \delta \varphi \rangle =-k_B\frac{\beta }{K_M},
     \nonumber \\
&& \langle \delta \beta  \delta \Phi \rangle 
    = k_B\frac{\beta \Phi}{C_Q},\ \ 
    \langle \delta \Phi\delta
    \Phi \rangle =-k_B\left(\frac{T}{K_M}+\frac{\Phi^2}
	 {C_Q}\right).
\end{eqnarray}
From the above, we can see that, when $b=2/(d-2)$, these second moments
are finite at the extremal limit. In this case, the extremal limit of
black hole is not a critical point, just as the $a=1$ dilaton black
holes in four dimensions \cite{Cai2}. Except this case, all second
moments diverge as the extremal limit is approached. This means that the
extremal limit is a critical point. Similar to the previous section, we 
can obtain the critical exponents in the `` first kind''
\begin{eqnarray}
&&\alpha=\psi=\gamma=\sigma=
\frac{4(d-3)^2+(d-4)(d-2)a^2}{(d-3)[2(d-3)+(d-2)a^2]},\nonumber \\
&& \beta=\delta
      ^{-1}=-\frac{2(d-3)^2-(d-2)a^2}{(d-3)[2(d-3)+(d-2)a^2]},
\end{eqnarray}
and in the ``second kind''
\begin{eqnarray}
&& \nu=\mu=\frac{2(d-3)^2-(d-2)a^2}{(d-3)[2(d-3)+(d-2)a^2]}, \nonumber
            \\
&& \eta=-\frac{(d-2)^2a^2}{2(d-3)^2-(d-2)a^2}.
\end{eqnarray}
From the scaling laws (\ref{scaling2}), we have the effective spatial 
dimension for an arbitrary dimensional dilatonic black hole
\begin{equation}
\label{d}
\bar{d}=\frac{(d-2)^2a^2}{2(d-3)^2-(d-2)a^2}.
\end{equation}
In general, the effective dimension will be on longer an integer. In
particular, when $a^2 > 2(d-3)^3/(d-2)$, the effective dimension is
negative. The implication has been discussed in \cite{Cai2}. This can 
be explained in the intersecting M-brane configurations \cite{Kle1}.
On the other hand, if $\bar{d}$ is remanded to be an integer, then the 
relation (\ref{d}) gives us a constrain on the coupling constant $a$.
To understand further its meaning is quite interesting.

\section{Conclusion}

 From the above and combining  with the results obtained 
 in Refs.\cite{Cai1,Cai2,Cai3}, we have the following 
 conclusions:
 
(1) The extremal limit of dilatonic and non-dilatonic black $p$-branes
 is critical point and corresponding critical exponents obey the
  scaling laws.  
  
(2) For the non-dilatonic black holes and black 
 strings, the effective spatial dimension is one. This result 
is also reached in the BTZ black holes and 3-dimensional black 
strings \cite{Cai1}. 

(3) For the non-dilatonic black $p$-branes (black 
string for $p=1$), the effective dimension is $p$, so does it for
the dilatonic black holes produced by the double-dimensional
 reduction of the non-dilatonic black $p$-branes.
  
(4) For other dilatonic black holes and  black $p$-branes, 
the effective spatial dimension  depends on the parameters 
in theories. 

(5) Furthermore, near the extremal limit of the
 non-dilatonic black $p$-branes, from 
 Eqs.~(\ref{temp1})-(\ref{adm}), we have
\begin{equation}
\label{relation}
S \sim T^{p} ,\ \ M-M_{\rm ext}\sim T^{p+1},
\end{equation}
where $M_{\rm ext}$ is the ADM mass of extremal black $p$-branes.
 Notice that the ADM mass and entropy of black $p$-branes are 
 extensive quantities with respect to the volume of $p$-branes.
 Thus, near the extremal limit, the thermodynamic properties 
 of non-dilatonic black $p$-branes can be described by the 
 blackbody radiation in ($1+p$) dimensions, which also further
 verify that  the effective spatial dimension is $p$. 
 For the 
 dilatonic black holes (\ref{solution}) with constant $a$ satisfying
 (\ref{parama}), the  equation (\ref{relation})
 is also valid. Although
 the entropy of dilatonic black holes is not an extensive quantity,
 the entropy can be regarded as the entropy density of the 
 non-dilatonic black $p$-branes. Thus it seems to imply that 
 these dilatonic black hole entropy can also be explained as 
 the way of $p$-branes, although the string coupling becomes 
 very large in this case.
 Therefore,  the entropy for the near extremal nondilatonic black holes, 
 black strings, and  black $p$-branes  may be explained by free 
 massless fields on the  world volume.

Recall the recent progress in understanding  entropy of 
black holes \cite{Horo},
 in which the constant dilaton field
 seems to be a necessary condition. Therefore, our conclusions 
are in complete agreement with the result of these 
investigations. In addition, Klebanov and Tseytlin \cite{kle} found 
that there are nondilatonic black $p$-branes whose near-extremal
entropy may be explained by free massless fields on the world volume. 
Thus  we also give an interpretation
 why the Bekenstein-Hawking entropy may be given a simple world 
 volume interpretation only for the non-dilatonic 
 $p$-branes (including the non-dilatonic black holes). 

\vspace*{1.cm}
This research was supported by the Center for Theoretical Physics of
Seoul National University.

\end{document}